\documentclass[aps,pre,amsmath,twocolumn,superscriptaddress]{revtex4-1}

\usepackage{graphicx}
\usepackage{color}
\usepackage[boxed]{algorithm2e}
\usepackage[caption=false]{subfig}

\begin{document}

\title{Greedy reduction of navigation time in random search processes}%

\author{Igor Trpevski}
\email[]{itrpevski@manu.edu.mk}
\affiliation{Macedonian Academy of Sciences and Arts}
\affiliation{IKT-Labs}

\author{Ljupco Kocarev}
\affiliation{Macedonian Academy of Sciences and Arts}
\affiliation{IKT-Labs}
\affiliation{Faculty for computer science and Engineering, St. Cyril and Methodius University}
\affiliation{BioCircuits Institute, UC San Diego, La Jolla, CA 92093-0402, USA}

\date{\today}

\begin{abstract}
Random search processes are instrumental in studying and understanding navigation properties of complex networks, food search strategies of animals, diffusion control of molecular processes in biological cells, and improving web search engines. An essential part of random search processes and their applications are various forms of (continuous or discrete time) random walk models. The efficiency of a random search strategy in complex networks is measured with the mean first passage time between two nodes or, more generally, with the mean first passage time between two subsets of the vertex set. In this paper we formulate a problem of adding a set of $k$ links between the two subsets of the vertex set that optimally reduce the mean first passage time between the sets.  We demonstrate that the mean first passage time between two sets is non-increasing and supermodular set function defined over the set of links between the two sets. This allows us to use two greedy algorithms that approximately solve the problem and we compare their performance against several standard link prediction algorithms. We find that the proposed greedy algorithms are better at choosing the links that reduce the navigation time between the two sets.
\end{abstract}


\maketitle

\section{Introduction}
The investigation of navigation and search processes on networks has a long tradition in the social sciences with a series of experiments conducted by Stanley Milgram \citep{milgram1967small} to confirm the existence of the postulated small-world phenomenon \citep{barabasi2002linked}. More recently, with development of the theory of complex networks, significant advances have been made in understanding these processes. Much of the studies  were concerned with the dependence of navigation on the topological characteristics of complex networks \citep{kleinberg2000navigation, hu2011possible, li2013optimal, perra2012random, de2014navigability}, ranking of nodes in terms of network navigation \citep{brin2012reprint, sanchez2012quantum} and the development of random search strategies \citep{noh2004random, cajueiro2009optimal, riascos2012long, di2015optimal}. 

With random search strategies a random walker starts from a source node and, in each step, according to a certain transition probability randomly moves to another node in the network  until it reaches a given target node. The details of the search strategies depend on the how the transition probability is defined. In order to measure the efficiency of a random search strategy one is typically interested in calculating the mean first passage time between two nodes. This can be computed by removing all out-links from the target nodes effectively transforming them into absorbing nodes. Recently, a new concept for studying absorbing random walks has been proposed \citep{mavro2015absorb}, where the average time to absorption is calculated between two \textit{sets of nodes}. One of the sets is called a set of query nodes, from which a random walker originates, and the other is a set of absorbing nodes where the path of the random walker finishes. The authors in \citep{mavro2015absorb} addressed the problem of finding a set of $k$ absorbing nodes, given a set of query nodes, for which a so called absorbing centrality quantity is minimal.

In this paper we formulate a related problem of finding the optimal placement of $k$ links that will reduce the average time to absorption between two sets of nodes in a graph. Formally, given a graph $G = (V,E)$, a set of query nodes $Q \subset V$ and a set of target nodes $C \subset V$ such that $Q \cap C = \emptyset$, we wish to add a set of $k$ links, originating from nodes in $Q$ and entering in nodes from $C$, that optimally reduce the expected time to absorption for a random walker starting on node in $Q$ and ending on a node in $C$. Exhaustively searching for the set of $k$ links which maximally reduce the expected time to absorption requires calculating this quantity for every possible combination of $k$ links out of the set of links between $Q$ and $C$. This represents a combinatorial problem with an exponential computational complexity which makes it computationally intractable even for small graphs, and one typically resorts to using heuristics which give approximate solutions. To solve this problem we prove that the average time to absorption between two sets of nodes is a supermodular non-increasing set function in terms of the set of edges between the nodes in $Q$ and $C$. This allows us to use greedy algorithms which yield an approximate solution to the optimal one. 

On the other hand, the fact that we need to predict a set of links allows us to frame the problem as a link-prediction problem and use many existing link-prediction algorithms as heuristics that approximately solve the problem \citep{lu2011link,lu2015toward}. We compare the Greedy algorithms with several link prediction algorithms and show that they produce better solutions in all the networks that have been considered.

The rest of the paper proceeds as follows. In section two we formally define the problem and  relate it to previous work. In section three we demonstrate the non-increasing and supermodularity properties of the function describing expected time to absorption. Then we propose two Greedy algorithms that exploit these two properties of the problem to solve it approximately. In section \ref{sec:results} we present the results of reducing the expected time to absorption between two sets of nodes along with the comparison of the Greedy algorithm and several other link prediction algorithms. We conclude the paper in section \ref{sec:conclusion}


\section{Problem statement}
\label{sec:prob-statement}
Let $G = (V,E)$ be a graph with a set of nodes $V$ and set of edges $E$, where the size of the graph is denoted by $N$, and the number of edges is denoted by $M$. Furthermore, we are given a partition of the graph $G$ into a set of query nodes $Q$ and a set of target nodes $C$ such that $Q \cap C = \emptyset$. The nodes in the set $C$ are transformed into absorbing nodes for a random walker originating in the set $Q$ by removing all their out-links. 

We define the expected number of steps for a random walk that starts from a query node $q \in Q$ until it gets absorbed in a node from $C$ as $m_{Q\to C}^q(E)$. The random walker starts on any of the query nodes with probability given by a starting distribution $\mathbf{s}(q)$. Without loss of generality we will consider only uniform distribution throughout the paper. To measure the expected time of absorption of a walker starting from any of the nodes in $Q$, we simply average over all possible query nodes:

\begin{equation} 
\label{eq-centrality}
m_{Q\to C}(E) = \sum_{q\in Q} \mathbf{s}(q) m_{Q\to C}^q(E)  
\end{equation}
In the case when there is a single node in the target set $C$ the quantity \ref{eq-centrality} reduces to the well-know random-walk centrality of a node \citep{noh2004random}.

The problem addressed in this paper is to place $k$ edges between the nodes in the sets $Q$ and $C$ so that the we maximally reduce the average time to absorption expressed with equation (\ref{eq-centrality}). This represents a combinatorial problem of choosing $k$ elements of the set of all possible links between $Q$ and $C$ that don't already exit in $E$. In the rest of the paper the index $Q\to C$ in equation (\ref{eq-centrality}) is dropped because we always consider random walks which originate in $Q$ and absorb in $C$. Note that the function (\ref{eq-centrality}) is similarly defined to the one given in \citep{mavro2015absorb} where it is defined in terms of the set of nodes, and the optimization problem is to find an optimal set of nodes. We now state some previous results from the theory on absorbing random walks.

\subsection{Absorbing random walks}

The absorbing random walk process is fully represented by a transition matrix $P$, where $P(i,j)$ expresses the probability that the random walker will move to node $j$ given that it is currently in node $i$. For an absorbing node $c$ the probability is given by the dirac delta function $P(c,j) = \delta_{cj}$. The set of query nodes $Q$ is also called a \textit{transient set}, because the random walk process will eventually leave this set. In the remainder of the paper we consider classical random walks but the results hold for more general models like random walks with restarts and random walks with teleportation. Formally if $N(i)$ is the set of neighbors of node $i\in Q$, and, $d_i$ its degree, the transition probabilities for a random walker to proceed from node $i$ are given by:

\begin{equation}
P(i,j) = \left\{
  \begin{array}{lr}
    1/d_i & \quad \text{if } j\in N(i) \\
    0 & \quad \text{otherwise}
  \end{array}
\right.
\label{eq:probability}
\end{equation}
The row-stochastic transition matrix of the random walk can be written in block form as:

\begin{equation}
 P = \left(
\begin{array}{cccc}
P_{QQ} & P_{QC} \\
0 & I \\
\end{array} %
\right)
\label{eq:transition}
\end{equation} 
where $P_{QQ}$ is a $|Q|\times|Q|$ sub-matrix that contains the probabilities of the set of transient nodes, and $P_{QC}$ is a $|C|\times|C|$ matrix that contains the transition probabilities from the transient to the absorbing set of nodes. The probability of the walk being at node $j$ at exactly $t$ time steps conditioned that it started at node $i$ is given by the $(i,j)$ entry of the matrix $P_{QQ}^t$. It can be demonstrated that the expected number of steps in which the random walk visits node $j$ given that it started in node $i$ is given by the $(i,j)$ entry of the \textit{fundamental matrix} of the absorbing random walk:
\begin{equation}
F = \sum_{t=0}^{\infty} P_{QQ}^t = (I-P_{QQ})^{-1}.
\label{eq:fundamental}
\end{equation}
The expected number of steps for a random walker starting on a query node $i$ and being absorbed into the set $C$ is given by the $i$-th element of the vector:
\begin{equation}
\mathbf{L}=\mathbf{L}_C = F\mathbf{1}
\end{equation}
where $\mathbf{1}$ is a column vector of length $Q$. Therefore the expected number of steps for a random walker starting on an arbitrary node in $Q$ until being absorbed in the set $C$ is given by averaging over all the query nodes:
\begin{equation} 
m(E) = \mathbf{s}^T\mathbf{L}_C = \mathbf{s}^T (I-P_{QQ})^{-1}\mathbf{1}
\label{eq-centrality-vectorised} 
\end{equation}

Obtaining the average time to absorption using equation (\ref{eq-centrality-vectorised}) involves calculating an inverse of a matrix, which in general involves $O(|Q|^3)$ calculations. Repeating this calculation for every combination of $k$ links is impossible with current computers. To overcome this problem we first show that the quantity (\ref{eq-centrality}) is a supermodular non-increasing set function, which allows us to use a greedy algorithm to obtain an approximate solution to the problem within a given approximation guarantee. Secondly, as we demonstrate later, each time we add a link we can use the Sherman-Morisson formula to update the inverse matrix in equation (\ref{eq-centrality-vectorised}), which requires $O(|Q|^2)$ computations. 

\section{Supermodularity of the problem and Greedy algorithms}
\label{sec:supermodularity}
Here, we state the lemmas for the two essential properties of the quantity (\ref{eq-centrality}) describing the expected time to absorption as a non-increasing and supermodular set function in terms of the set of links between the two sets $Q$ and $C$. The proofs of the lemmas are given in the appendix. 

\textbf{Lemma 1}. For all subsets $X \subset Y \subset E_{QC}$ it holds that $m(Y) \leq m(X)$, where $E_{QC}$ is the set of all links between $Q$ and $C$. 

\textbf{Lemma 2} For all subsets $X \subset Y \subset E_{QC}$ and $e \notin Y$ where, it holds that:
\begin{equation*}
m(X) - m(X\cup \{e\}) \geq  m(Y) - m(Y\cup \{e\}). 
\label{eq:supermodularity}
\end{equation*}
such that $e$ is an edge from $Q$ to $C$. 

Lemma 1 tells us that whenever we add a link between the sets $Q$ and $C$ the quantity (\ref{eq-centrality}) will either decrease or stay the same. Lemma 2, states that, given a set $Y$ and one subset of it $X$, adding a new link $e$ to $X$ yields a greater decrease in (\ref{eq-centrality}) than adding it to $Y$. This property of set functions is analogous to the concavity property of ordinary functions.

It is well established that submodular and supermodular problems can be approximatelly solved by greedy algorithms \citep{kempe2003maximizing, sviridenko2015optimal}. Here we present two variants of a greedy algorithm that approximatelly solve the problem of adding the $k$ edges that optimally reduce the expected time to absorption between the query set $Q$ and target set $C$. These $k$ links are chosen from the set of candidate links $E'$ between the $Q$ and $C$ that are not already present in the graph.

\begin{figure}
\includegraphics[trim={0 0 4cm 0},clip]{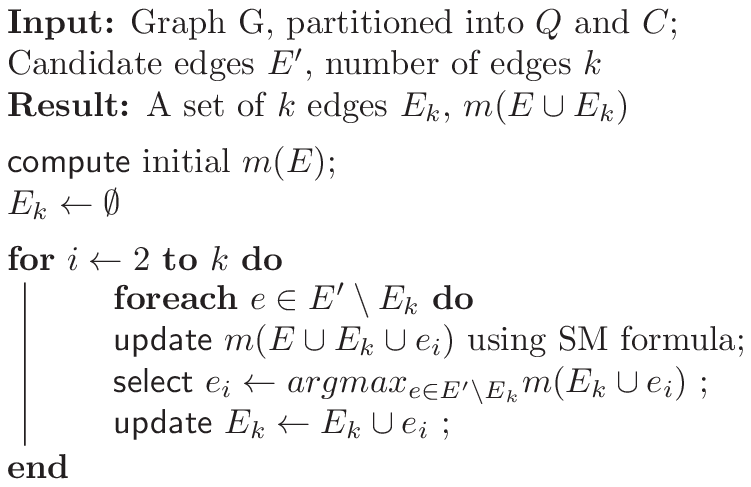}
\caption{Greedy descent algorithm}
\label{fig:algo1}
\end{figure}

We depict the first algorithm, also called greedy descent algorithm in figure \ref{fig:algo1}. It starts with a graph partitioned into a set $Q$ and a set $C$, as well as the set of candidate edges $E'$. In each step $i$, we add a link $e_i \in E'$ that maximally reduces the quantity (\ref{eq-centrality}). For every candidate link, this requires the calculation of expression (\ref{eq-centrality-vectorised}) which involves inverting a matrix in $O(N^3)$ steps, or in the case of the fundamental matrix $O(|Q|^3)$. If the fundamental matrix $F$ is already calculated, adding a single link can be seen as a perturbation or a rank-1 update to the matrix $F$, which can be done in $O(|Q|^2)$ steps using the Sherman-Morrison formula. In order to add the out-link from $Q$ to $C$ which maximally reduces the quantity (\ref{eq-centrality}) in every step, we need to apply the Sherman-Morison step $|Q|$ times. This leads to a total running time of $k|Q|^3$ for the greedy algorithm. 

\begin{figure}
\includegraphics[trim={0 0 4cm 0},clip]{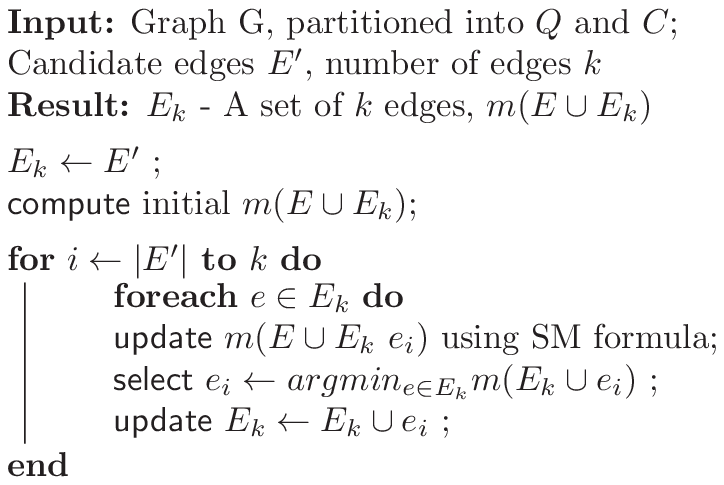}
\caption{Reverse greedy algorithm}
\label{fig:algo2}
\end{figure}

The second algorithm, depicted in figure \ref{fig:algo2} is known as reverse greedy descent and starts by first adding all the edges from the set $E'$ which, according to Lemma 1, yields a minimal value for the quantity (\ref{eq-centrality}). Then, in each iteration it removes the 'worst edge' i.e. the edge that minimally increases the expected time to absorption (\ref{eq-centrality}). It stops when it comes to the number of edges $k$ that we wished to add to the original set of links. The remaining $k$ links is the approximate solution to the problem. This algorithm has a running time of  $(|E'|-k)|Q|^3$, which is larger than the running time of algorithm 1, when our aim is to add a small number of links but it is more efficient when a large number of links need to be added (larger than $|E'|/2$.  Another reason we also show the reverse greedy algorithm is that it has a proven approximation guarantee to the optimal solution of the problem, while this in general is not known for the typical greedy descent algorithm \citep{il2001approximation}, although under certain constraints on the optimization function there is an approximation guarantee to the solution yielded by algorithm 1 \citep{sviridenko2015optimal}. 

\section{Results}
\label{sec:results}

\begin{figure*}[htp]
\label{small}
\subfloat[Zachary Karate club $|C|=3$]{%
  \includegraphics[trim=0 0 0 1.8cm,clip,width=0.33\textwidth]{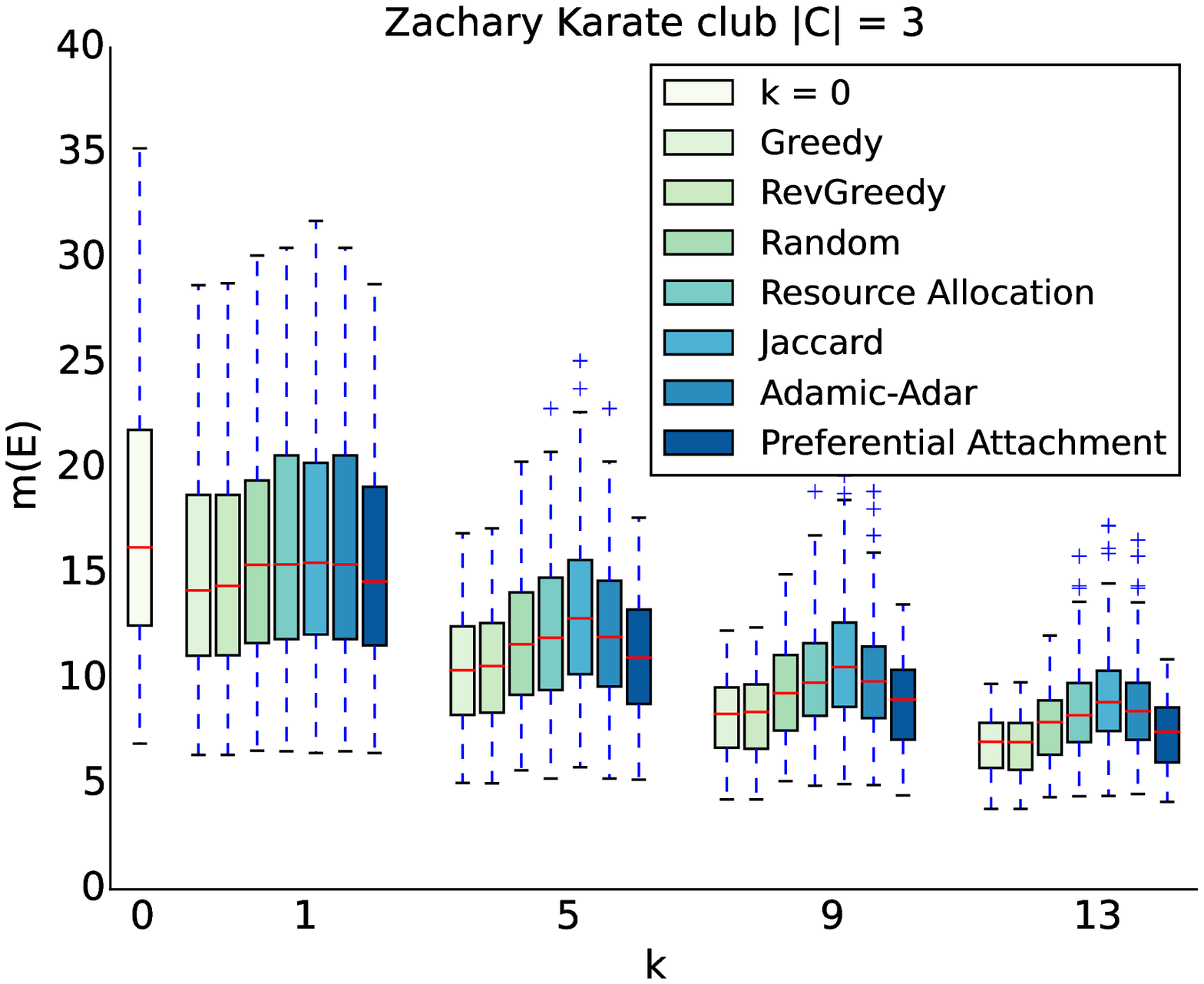}%
}
\subfloat[Dolphins $|C|=3$]{%
  \includegraphics[trim=0 0 0 1.8cm,clip,width=0.33\textwidth]{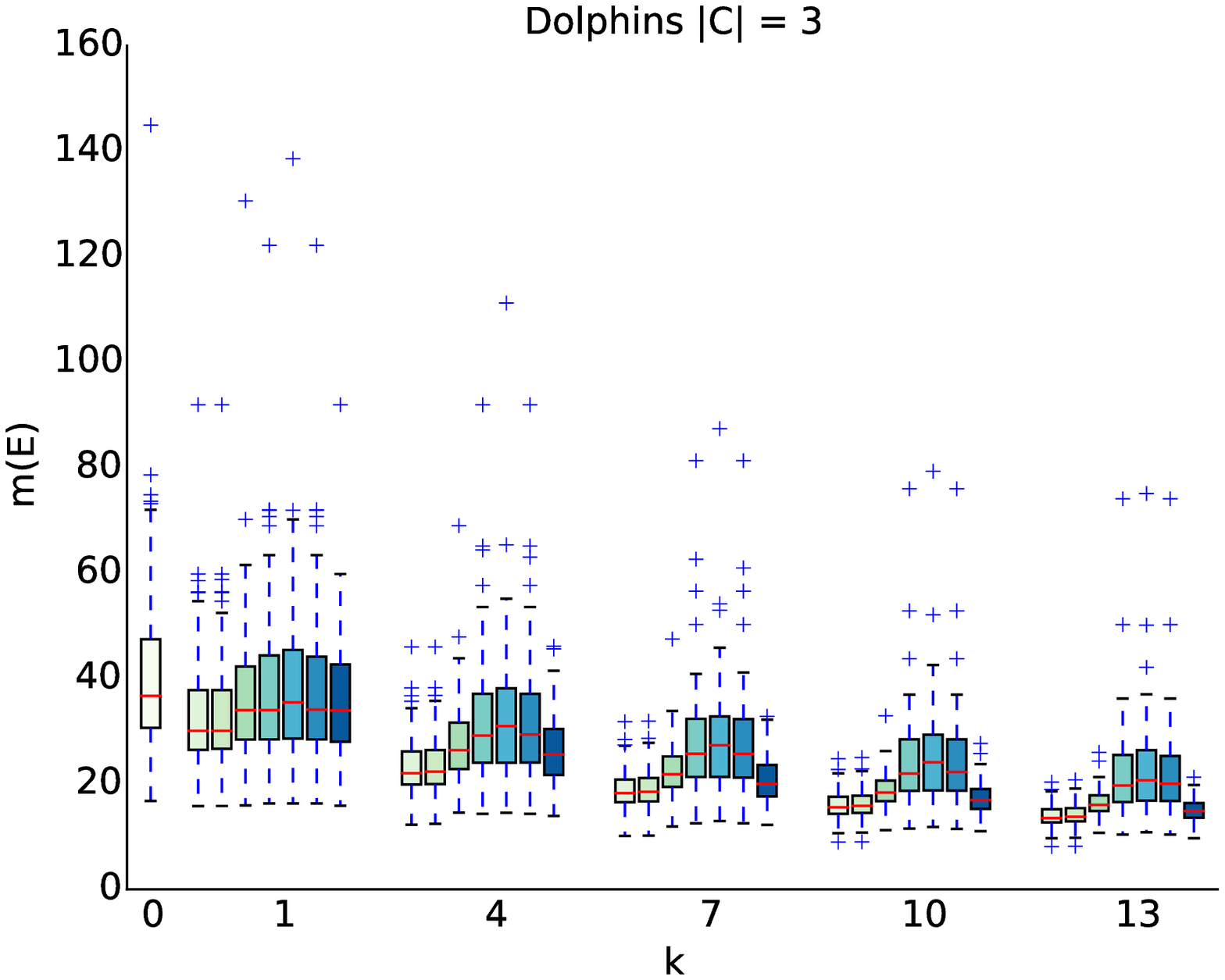}%
}
\subfloat[Social patterns $|C|=10$]{%
  \includegraphics[trim=0 0 0 1.8cm,clip,width=0.33\textwidth]{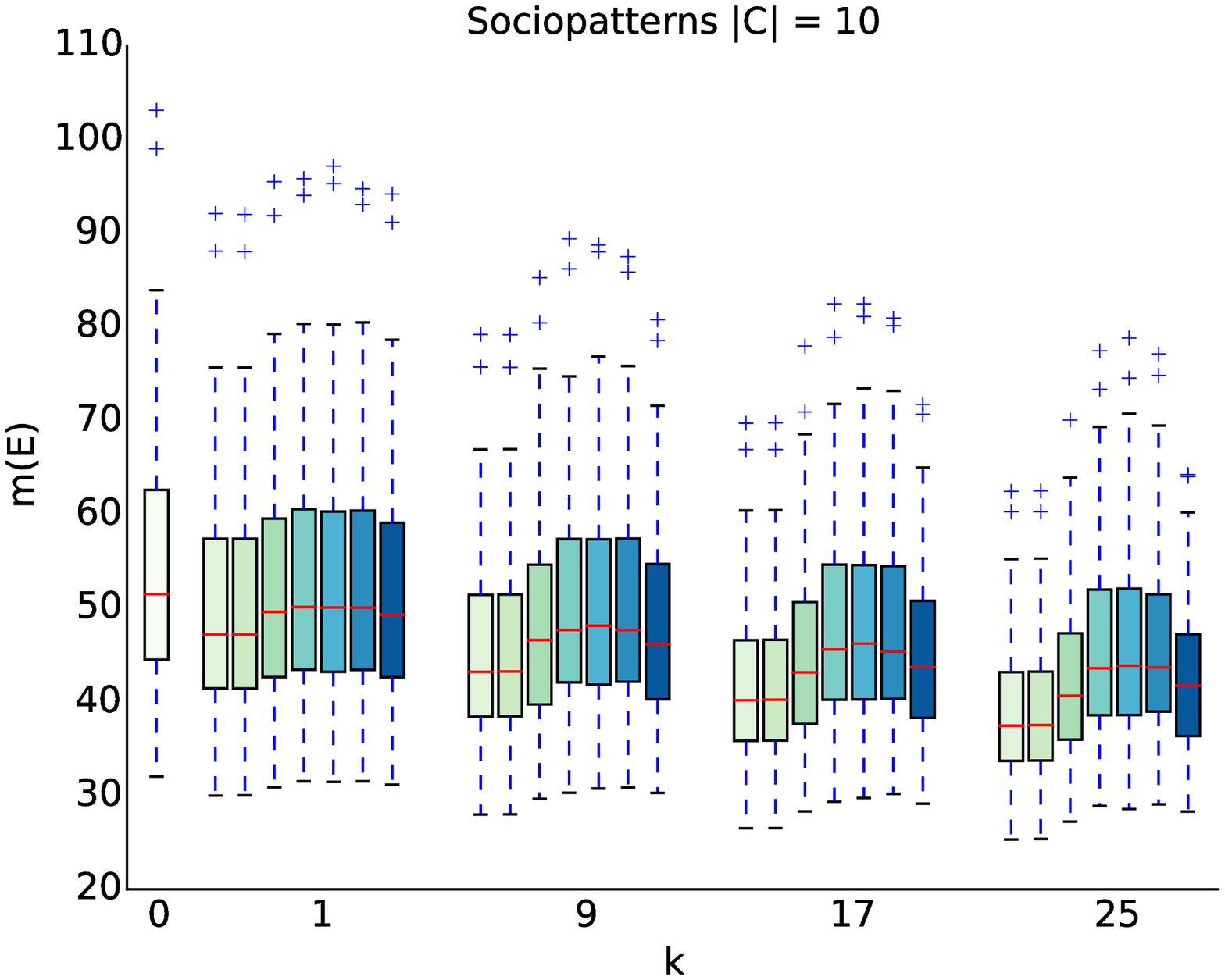}%
}
\caption{(Color online) The expected time to absorption between sets $Q$ and $C$ on three small real-world graphs. The performance of the greedy algorithms is evidently better than the standard link prediction algorithms, of which only the preferential attachment algorithm offers better than random performance.}

\end{figure*}

We compare the reduction in expected time to absorption obtained with the two greedy algorithms against several standard link-prediction algorithms: resource allocation \citep{zhou2009predicting}, Jacquard coefficient, Adamic-Adar index and preferential attachment index \citep{liben2007link}. For each of this algorithms we calculate the scores over the set of possible candidate links between the query set $Q$ and the absorbtion set $C$, and pick the $k$ highest ranking links. Subsequently, after adding these links we measure the reduction in the expected time to absorption. We also provide results for a uniform random choice of $k$ links. These algorithms have much better running times than the greedy algorithms presented in this paper. In order to make the results reproducible we made the code available at \citep{github}.

For every graph $G$ we randomly generate a partition into a query set $Q$ and target set $C$, so that the query set is a connected subgraph. We repeat this process $L$ times and for each partition we calculate the reduction in expected time to absorption after adding $k$ links. In figure 3 we show the results for three small graphs: the Zachary karate klub graph \citep{konect:ucidata-zachary}, the network of interactions between Dolphins \citep{konect:dolphins} and a network of face-to-face behavior of people during an exhibition \citep{konect:sociopatterns}. Both greedy algorithms perform better than the other link-prediction algorithms, and in many cases the solutions they yield are very close. From the latter group, only the preferential attachment index compares with a baseline random strategy of adding links between $Q$ and $C$. In terms of a random- navigation strategy this implies that, in some cases, adding links between highly connected nodes yields no better reduction in navigation time than randomly adding links. The outcome arguably  depends on the initial topological characteristics of the graph such as the degree distribution and assortativity coefficient.

In figure 4 we show similar results on larger graphs of the size $O(N^3)$: a protein interaction network of the yeast organism (YPI) \citep{konect:coulomb2005}, a network of european roads \citep{konect:eroads} and an email communication network at the University of University Rovira i Virgili \citep{konect:guimera03}. The reverse greedy algorithm runned very slow on these networks and no results are shown for it, although we expect similar results to those of the greedy descent algorithm. On the YPI and road networks the greedy algorithm produced a significant reduction in average time to absorbtion outperforming all other algorithms, with very little spread over the different partitions. The preferential attachment algorithm outperforms the random strategy on the YPI network which has a negative assortativity coefficient of -0.16, while the random strategy is better on the E-road network which has a positive assortativity coefficient of 0.12. Based on this it can be argued that adding links between highly connected nodes to reduce the negative assortativity is good in terms of random navigation between two sets of nodes. The opposite might also be argued as well - on a network with positive assortativity randomly adding links between the two sets is a better strategy to achieve reduction in navigation time.   The email communication network is an example where (\ref{eq-centrality}) is relatively flat function and no algorithms produced significant reduction in average time to absorption.

\begin{figure*}[htp]
\label{large}

\subfloat[Yeast protein interaction $|C|=5$]{%
  \includegraphics[trim=0 0 0 1.8cm,clip,width=0.33\textwidth]{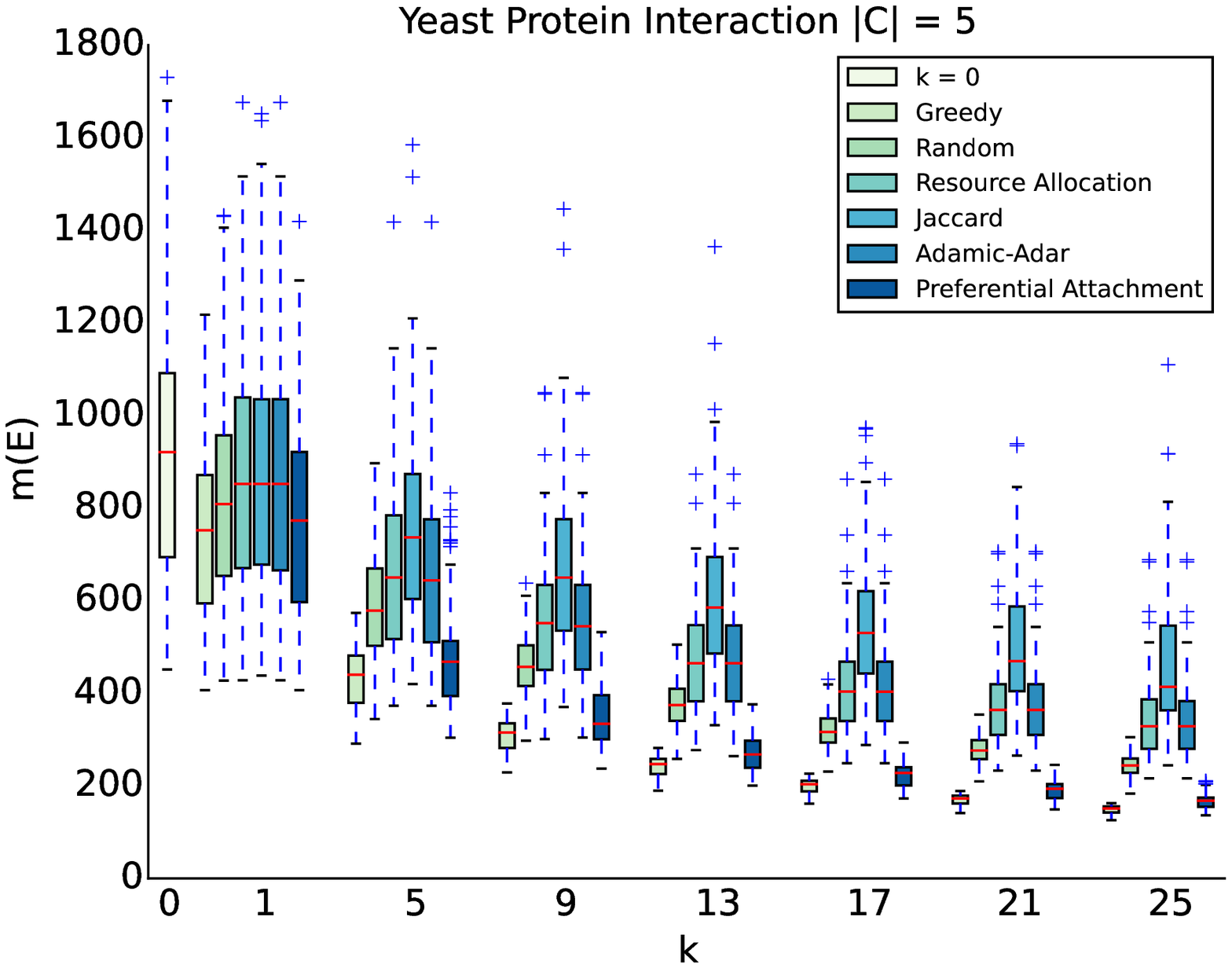}%
  }
\subfloat[European E-road $|C|=5$]{%
  \includegraphics[trim=0 0 0 1.8cm,clip,width=0.33\textwidth]{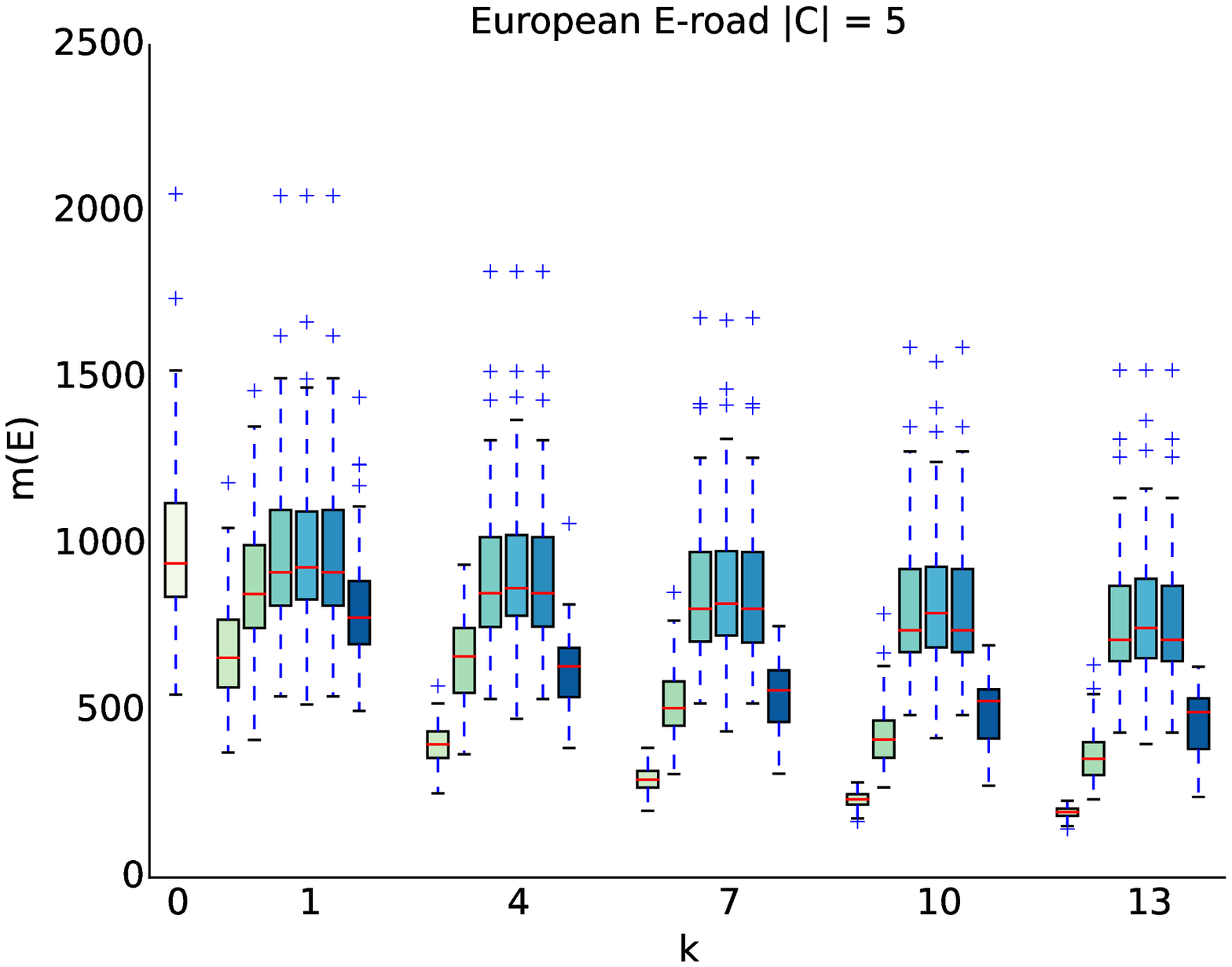}%
}
\subfloat[Email of Rovira I Virgili $|C|=5$]{%
  \includegraphics[trim=0 0 0 1.8cm,clip,width=0.33\textwidth]{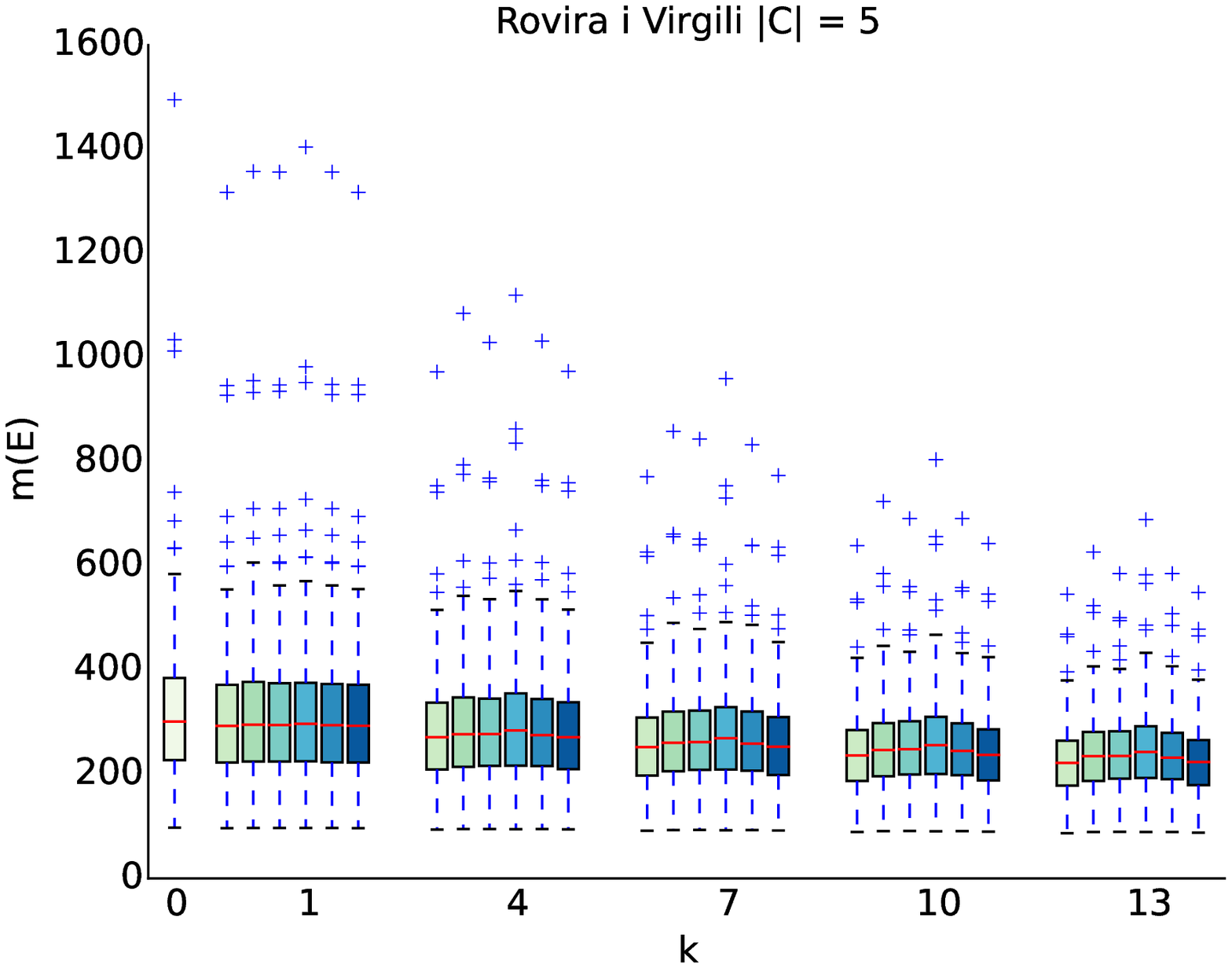}%
}
\caption{(Color online) Reduction in expected time to absorption on three larger real-world graphs. The Greedy algorithm outperforms the other link-prediction algorithms in the cases a) and b) while no algorithm manages to significantly reduce navigation time in the case c)}

\end{figure*}

\section{Conclusion}
\label{sec:conclusion}

We formulate the problem of finding the optimal placement of $k$ links that will reduce the expected time to absorption between two sets of nodes in a graph for a random walk process. We proposed the two greedy heuristics that approximately solve the problem exploiting the supermodularity property of the quantity (\ref{eq-centrality}). The problem can be equivalently formulated in terms of finding a set of existing $k$ links between the sets $Q$ and $C$ that are most important for navigation. Since the problem can be considered as a link-prediction problem we compared the Greedy algorithms with several link prediction algorithms and showed that they yield better solutions in all the networks that have been considered. 

\appendix
\section{}

We first show that (\ref{eq-centrality}) is a non-increasing set function. Let $E_{QC}$ be the set of edges originating from the query nodes $Q$ and entering the absorbing nodes in $C$.

\textbf{Lemma 1}. For all subsets $X \subset Y \subset E_{QC}$ it holds that $m(Y) \leq m(X)$. 

\textit{Proof.} It is sufficient to prove monotonicity only for an arbitrary term in the summation (\ref{eq-centrality}). Let $q$ be an arbitrary node from $Q$ with one or more edges towards nodes from the set $C$ and denote this set of edges as $Y_q \subset N(q)$. Let $Y \subset E_{QC}$ be a set that includes these edges (i.e. $Y_q \subset Y$) and let $X = Y\setminus Y_q$. The proof can easily be generalized to the case when some of the edges in $Y_q$ belong to $X$ as well. 

Now consider all paths of a random walk process which pass through the node $q$ and are absorbed through the set of edges $Y$. We can divide them in two classes: i) paths that are absorbed through the set of edges $X$, denoted with $\mathcal{P}(X)$ and, ii) paths that are absorbed through the set of edges $Y_q$, denoted with $\mathcal{P}(Y_q)$. 
For a path $p$ we denote with $Pr[p]$ the probability of its occurrence and with $l[p]$ its length. 
For the two disjunct sets of paths $\mathcal{P}(X)$ and $\mathcal{P}(Y_q)$ we can calculate the expected number of steps for a walk that is found at $q \in Q$ until it gets absorbed through the set of edges $Y$. Given that the random walk has reached node $q$ it will continue with probability $1-|Y_q|/d_q$ to a neighbor of $q$ in $Q$ which we denote with $r \in N_Q(q)$, or will get absorbed with probability $|Y_q|/d_q$. Thus the expected number of steps for all paths that arrived at $q$ and are absorbed through the edges in $Y$ is:

\begin{eqnarray}
m^q(Y) & = & \sum_{p\in \mathcal{P}(Y)} Pr[p] l[p] \nonumber \\
& = & E[l_q] + \frac{|Y_q|}{d_q} + \left(1-\frac{|Y_q|}{d_q)}\right)  \left(1+m^{r} (Y) \right) \nonumber \\
& = & E[l_q] + 1 + \left(1-\frac{|Y_q|}{d_q)}\right) \left(m^{r} (Y) \right)
\label{eq:abs-y}
\end{eqnarray}
where $m^{r}(Y)$ is the expected number of steps until absorption for a random walk that continued at any of the neighbors of $q$ in $Q$, and $ E[l_q]$ is the expected number of steps that the random walker has made before it visited $q$. 

We can write a similar expression for the expected number of steps until absorption of a random walk which visits $Q$ and gets absorbed through the set of links $X$. 

\begin{equation}
m^q(X) =  \sum_{p\in \mathcal{P}(X)} Pr[p] l[p] = E[l_q] + 1 + m^{r} (Y)
\label{eq:abs-x}
\end{equation}
It is straightforward to see from the expressions (\ref{eq:abs-y}) and (\ref{eq:abs-x}) that $m^q(Y) \leq m^q(X)$. The equality holds for the case when the node $q$ doesn't have any links towards the set of absorbing nodes $C$. 

\textbf{Lemma 2} For all subsets $X \subset Y \subset E_{QC}$ and $e \notin Y$ it holds that 
\begin{equation}
m(X) - m(X\cup \{e\}) \geq  m(Y) - m(Y\cup \{e\}).
\label{eq:supermodularity-appendix}
\end{equation}
such that $e$ is an edge from $Q$ to $C$.

\textit{Proof.} As in the previous proof it is sufficient to demonstrate that the inequality holds for an arbitrary node $q$ from the set of query nodes $Q$. Using the same notation as with the proof of the monotonicity property we can write the four terms in (\ref{eq:supermodularity-appendix}) as:
\begin{eqnarray}
m^q(X) & = & E[l_q] + 1 + m^{r}(X) \nonumber \\ 
m^q(X \cup \{e\}) & = & E[l_q] + 1/d_q + \frac{d_q-1}{d_q}(1 + m^{r}(X)) \nonumber \\
m^q(Y) & = & E[l_q] + 1 + m^{r}(Y) \nonumber \\
m^q(Y \cup \{e\}) & = & E[l_q] + 1/d_q + \frac{d_q-1}{d_q}(1+m^{r}(Y)) \nonumber \\
\end{eqnarray}
and applying these into equation (\ref{eq:supermodularity-appendix}) we obtain:

\begin{eqnarray}
m^{r}(X) - \frac{d_q-1}{d_q}(1+m^{r}(X)) & \geq \nonumber \\ m^{r}(Y) - \frac{d_q-1}{d_q}(1+m^{r}(Y)) \nonumber
\end{eqnarray}

or equivalently, $m^{r}(X)\geq m^{r}(Y)$, which we now already that is true from Lemma 1.


\bibliography{optimizing_centrality}

\end{document}